\documentclass[12pt]{article}

  \usepackage{graphicx}
  \usepackage{epsfig}
  \usepackage{amssymb}
  \usepackage{subfigure}
  \usepackage{amsmath} 

\evensidemargin - .5truecm
\allowdisplaybreaks[1]

\def \gsim
{\raisebox{-3pt}{$\>\stackrel{>}{\scriptstyle\sim}\>$}}

\begin{document}

\newcommand{\CC}{{\mathbb C}}
\newcommand{\RR}{{\mathbb R}}
\newcommand{\ZZ}{{\mathbb Z}}
\newcommand{\QQ}{{\mathbb Q}}
\newcommand{\NN}{{\mathbb N}}
\newcommand{\beq}{\begin{equation}}
\newcommand{\eeq}{\end{equation}}
\newcommand{\beal}{\begin{align}}
\newcommand{\eeal}{\end{align}}
\newcommand{\nn}{\nonumber}
\newcommand{\bea}{\begin{eqnarray}}
\newcommand{\eea}{\end{eqnarray}}
\newcommand{\ba}{\begin{array}}
\newcommand{\ea}{\end{array}}
\newcommand{\bfig}{\begin{figure}}
\newcommand{\efig}{\end{figure}}
\newcommand{\bc}{\begin{center}}
\newcommand{\ec}{\end{center}}

\newenvironment{appendletterA} 
{
  \typeout{ Starting Appendix \thesection }
  \setcounter{section}{0}
  \setcounter{equation}{0}
  \renewcommand{\theequation}{A\arabic{equation}}
 }{
  \typeout{Appendix done}
 }
\newenvironment{appendletterB}
 {
  \typeout{ Starting Appendix \thesection }
  \setcounter{equation}{0}
  \renewcommand{\theequation}{B\arabic{equation}}
 }{
  \typeout{Appendix done}
 }

%
%
%
%

\begin{titlepage}
\nopagebreak

\renewcommand{\thefootnote}{\fnsymbol{footnote}}
\vskip 2cm
\begin{center}
\boldmath

{\Large\bf $1/n$ Expansion of Resonant States}

\unboldmath
\vskip 1.cm
{\large  U.G.~Aglietti}
\vskip .3cm
{\it Dipartimento di Fisica, Universit\`a di Roma ``La Sapienza'' } 
\end{center}
\vskip 0.7cm

\begin{abstract}  
We present a general analytic expansion in powers of $1/n$ of the resonant states
of quantum-mechanical systems, where $n=1,2,3,\cdots$ is the 
excitation number.
Explicit formulas are obtained for some potential barrier models.

 
\end{abstract} 
\vfill 
\end{titlepage}     
 
\setcounter{footnote}{0} 
\newpage 

\begin{figure}[ht]
\begin{center}
\includegraphics[width=0.5\textwidth]{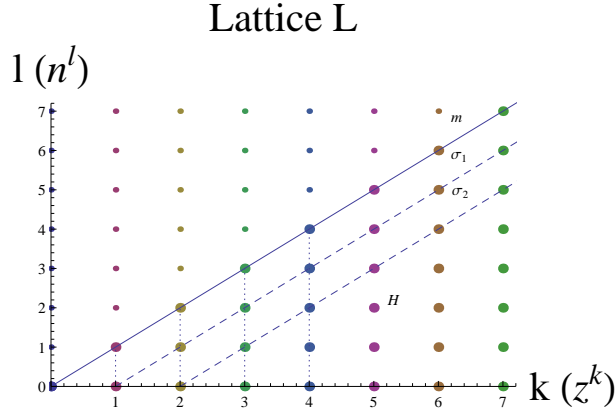}
\footnotesize
\caption{
\label{figura1}
\it The (big) points $p=(k,l)$ with $k=0,1,2,3,\cdots$ and $0 \le l \le k$
on the lattice $L$ represent the terms $z^k n^l$ in the expansion of 
the observable $O(n,z)$ (see text).
The (dotted) vertical segments represent the terms associated to some power of $z$.
The (continuous) main diagonal $m$, of equation $l=k\ge 0$, contains the points 
corresponding to the resummed terms in $\sigma_0$. 
The two (dashed) half-lines parallel to $m$ correspond to the functions $\sigma_1$
and $\sigma_2$. 
}
\end{center}
\end{figure}

A resonance in quantum mechanics originates from the coupling of a
state in the discrete spectrum to a continuum \cite{gamow,refbase1,refbase2}, such coupling
generally increasing with the energy of the resonance.
As well known indeed, decay widths increase with the resonance energy, the reason for that being
kinematic: by increasing the energy, the available phase space to the decay products becomes larger.
The idea, roughly speaking, is to take advantage of this fact to generate an expansion for 
high-energy resonances, which actually works also for low-energy ones.
Let us consider a Hamiltonian $\hat{H}$ depending, for simplicity's sake, on a single 
coupling $z$,
\beq
\hat{H} \, = \, \hat{H}(z),
\eeq
and possessing an infinite tower of resonances $\psi_n(z)$, $n=1,2,3,\cdots$ going, in the free limit $z\to 0$,  
into the discrete spectrum of the limiting Hamiltonian
\beq
\hat{H}_0 \,\, \equiv \,\,  \lim_{z\to 0} \, \hat{H}(z) .
\eeq
As example of a free Hamiltonian, one may consider for example a particle in a box or a harmonic
oscillator.
Conversely, we may consider a Hamiltonian $\hat{H}_0$ possessing an infinite discrete spectrum
$\left\{\phi_n\right\}$, and add to it an interaction Hamiltonian involving a small coupling $z$,
\beq
\hat{H}_0 \,\,\, \to \,\,\, \hat{H}(z) \, \equiv \, \hat{H}_0 \, + \, \hat{H}_I(z) .
\eeq
For a generic interaction, by switching on the coupling from $z=0$ to $z\ne 0$ ($|z|\ll 1$ always),
the discrete states $\phi_n$ disappear from the spectrum and, according to the adiabatic continuity
principle \cite{anderson}, go one-to-one into long-lived resonance states.
As we are going to explicitly show in the following examples, an observable $O$ related to the
$n^\mathrm{th}$ resonance, has a general double series expansion of the form:
\beq 
\label{double}
O(n,z) \, = \,\alpha  n +  \sum_{k=0}^\infty \sum_{l=0}^k c_{k,l} n^l z^k =
\alpha  n + c_{0,0} + c_{1,1} n z +  c_{1,0} z + c_{2,2} n^2 z^2 + c_{2,1} n z^2 + c_{2,0} z^2 + \cdots ,
\eeq
where $\alpha$ and $c_{k,\,l}$ are complex coefficients.
By collecting together all the terms with the same power of $z$, one can rewrite the above formula as
an ordinary series:
\beq 
\label{double2}
O(n,z) \, = \, \alpha  \, n \, + \, \sum_{k=0}^\infty p_k(n) \, z^k , 
\eeq
where $p_k(n)$ is the polynomial in $n$ of degree $k$,
\beq  
p_k(n) \, \equiv \, \sum_{l=0}^k c_{k,\,l} \, n^l .
\eeq
In the free limit at a fixed resonance,
\beq
\label{standard}
z \, \to \, 0, \,\,\,\,\,\,\,\,    \mathrm{fixed} \,\, n ,
\eeq
we obtain for the observable the free value:
\beq
O(n,z=0) \, = \, \alpha  \, n \, + \, c_{0,0} \, ,
\eeq
whose first term has been separated out of the above series.
The standard perturbative expansion involves the truncation
of the series in $z$ at some (finite) order, let's say $k=N$, 
together of course with the exact evaluation of the corresponding 
coefficients:
\beq
\label{perturbative}
O(n,z) \, \simeq \, \alpha  \, n \, + \, \sum_{k=0}^N p_k(n) \, z^k .
\eeq
Let us notice that, in the limit specified in eq.(\ref{standard}),
one is perturbing a {\it free} model, in which resonances are stable,
as there is no coupling of the latter with the continuum.
Now, the key observation is simply that, in eq.(\ref{double}), 
each power $k$ of the coupling $z$ is multiplied 
by a smaller or equal power $l$ of the excitation number $n$,
\beq
\label{landk}
n^l \, z^k : \,\,\,\,\,\,\,\,\, 0 \, \le \, l \, \le \, k < \, \infty .
\eeq
Let us then assume we are interested in the study of a high-energy resonance,
\beq
\label{nlarge}
n \, \gg \, 1,
\eeq
in the weakly-interacting domain,
\beq
0 \, < \, |z| \, \ll \, 1,
\eeq
such that the product of the above variables is of order one:
\beq
n \, z \, = \, \mathcal{O}(1) .
\eeq
Because of the inequality (\ref{nlarge}), to a first approximation, the double series in eq.(\ref{double}) 
can be replaced by a single series with the highest possible power of $n$ for each power of $z$, i.e. 
with $l \, = \, k$. Eq.(\ref{double}) then simplifies to:
\beq
\label{rough}
O(n,z) \, \approx \,  \alpha  \, n \, + \, \sum_{k=0}^\infty c_{k,\,k} \, (n z)^k
\, = \, \alpha  n + c_{0,0} + c_{1,1} n z + c_{2,2} (n z)^2 + c_{3,3} (n z)^3 + \cdots .
\eeq
In order for the above approximation to be sensible, one has to assume that the coefficients
$c_{k,l}$ do not behave in a wild way with $l$: that is {\it a priori} possible, but it turns
out to be false {\it a posteriori} in all the examples (see later).
The crucial point is however that the "effective coupling" of the $n^\mathrm{th}$ resonance $\psi_n$ 
to the continuum, is actually $n \, z$, rather than $z$. 
In other words, not one, but two different couplings are involved in resonance phenomena: 
a state-independent coupling $z$, entering $\hat{H}$, 
and an effective, state-dependent coupling $n z$, which controls the actual size of
the coupling of the $n^\mathrm{th}$ resonance to the continuum.
The physical situation described above formally corresponds 
to the correlated limit:
\beq 
\label{largen}
n \, \to \, \infty , \,\,\,\,\,\,\, z \, \to \, 0 ,
\eeq
with the variable product approaching a non-zero constant:
\beq
\label{correlate}
n \, z \,\, \to \,\, \mathrm{const} \, \ne \, 0, \, \infty.
\eeq
Let us stress that the limit (\ref{largen}-\ref{correlate}) involves a delicate
balance of effects: If we send $z\to 0$ without increasing $n$, metastable dynamics
disappear from the model, because resonances become stable states,
i.e. eigenfunctions of $\hat{H}$.
On the other side, if we send $n\to\infty$ by keeping $z$ constant (and not zero), 
resonances become so wide as to loose some meaning and their description
becomes less accurate (see eq.(\ref{Gamma0})).
The next step is to notice that eq.(\ref{rough}) can be systematically improved,
by thinking to its right-hand-side as the lowest-order term of the following function
series: 
\beq 
\label{exact}
O(n,z) \, = \, \alpha  \, n \, + \, \sum_{k=0}^\infty \frac{1}{n^k} \, \sigma_k(\zeta) 
\, = \, \alpha  \, n \, + \, \sigma_0(\zeta) + \frac{1}{n} \, \sigma_1(\zeta)
+ \frac{1}{n^2} \, \sigma_2(\zeta) + \cdots ,
\eeq
where:
\beq
\sigma_k(\zeta) \, \equiv \, \sum_{l=k}^\infty c_{l,\, l-k} \, \zeta^l ,
\,\,\,\,\,\,\,\, k \ge 0 .  
\eeq
We have defined the $n^\mathrm{th}$ resonance coupling as $\zeta \, \equiv \, n \, z$.
The following remarks are in order.
Eq.(\ref{exact}) is exact but, as it stands, is just a rearrangement of the
$z$-expansion in eq.(\ref{double2}). The general problem of analytic computations 
of resonances is that, apart from exceptional cases, one is not 
able to exactly resum the series for $O(n,z)$ in {\it any} form, so is forced 
to use truncated formulas.
In our new scheme, we then truncate the function series above to some $k=K<\infty$:
\beq 
\label{approx}
O(n,z) \, \cong \, \alpha  \, n \, + \, \sum_{k=0}^K \frac{1}{n^k} \, \sigma_k(\zeta) .
\eeq
As we will see later, this equation effectively allows for simple analytic results in various models.
Roughly speaking, the idea of the $1/n$-expansion is that eq.(\ref{approx}) provides
a better approximation to the exact theory than the standard perturbative expansion,
eq.(\ref{perturbative}). To begin with, the $1/n$-expansion is obviously better, 
by construction, at least for high-energy resonances. 
Because of eq.(\ref{landk}), {\it any} truncated expansion in $z$,
at whatever order $N$, cannot describe resonances with excitation number
\beq
n \, \gsim \, n_{cr} \, \equiv \, \frac{1}{|z|} , 
\eeq
as $|n z| \gsim 1$ in this case, 
implying that the neglected terms are of the same size as the computed ones
($|z|\ll 1$, $z\ne 0$ and fixed: "fixed physics"). 
The critical value of the excitation number, $n_{cr} = n_{cr}(z)$,
represents then an "essential barrier" for fixed-order perturbation theory. 
In order to have a graphical representation of the $z$ and the $1/n$ expansions, 
let us consider a square plane lattice $L$, consisting of all points with
positive integer coordinates (see fig.\ref{figura1}),  
\beq
L \, \equiv \, \Big\{ (k,l); \,\, k,l=0,1,2,3,\cdots \Big\} .
\eeq
We select the points $p=(k,l)$ representing the terms $c_{k,\,l} z^k n^l$ in 
eq.(\ref{double}) lying, according to this equation, below or on the top of
the half-line $m$ of equation $l=k \ge 0$, namely the main diagonal. 
Geometrically, these points $p$ fill the lower half $H$ of $L$.
Each perturbative order, i.e. each power $z^s$, 
corresponds to the vertical segment $k=s$, with $0 \le l \le s$.
The points corresponding to $\sigma_0$ lie instead on the main diagonal $m$.
In general, each function $\sigma_s(\zeta)$ corresponds to the half-line
of equation $l=k-s$ with $k \ge s$, parallel to and below the main diagonal $m$. 
By means of a (truncated) $z$-expansion, one then fills a triangle
inside $H$ with one vertex at the origin, while with the $1/n$ expansion
one fills a tilted infinite strip inside $H$, with the upper side on the diagonal $m$.
In brief: with the perturbative expansion, one sums the double series of the observable
$O(n,z)$ along vertical segments, while with the $1/n$ expansion one sums along diagonals. 
Let us notice that at the lowest (non-trivial) order in $1/n$, we obtain a different
limiting model compared to the $z$-expansion, which we may just call the "$n$$=$$\infty$ model". 
This model already contains metastable states, described by the leading-order
function $\sigma_0(\zeta)$. One is then perturbing this model
by taking $n$ large but finite, obtaining corrections proportional to powers of $1/n$,
of the form $\sigma_k(\zeta)/n^k$ according to eq.(\ref{approx}).
Contrary to the usual perturbative expansion in $z$ in eq.(\ref{perturbative}),
the $1/n$-expansion in eq.(\ref{approx}) involves an approximate resummation of 
the perturbative series to all orders, as each term $\sigma_k(\zeta)$ involves the summation
of infinitely many terms in $z$.
Therefore, in this sense, we may say that a (truncated) $1/n$ expansion is 
non-perturbative in character because, even though it is not exact, it allows 
to describe some phenomena which are unreachable or invisible to a truncated expansion in $z$.
It is clear that the expansions in $z$ and in $1/n$ of the same model have 
to be mutually consistent. That implies in particular that, by expanding
the $\sigma_k(\zeta)$'s in powers of $\zeta$, one should obtain the
standard perturbative expansion.
On the contrary, by means of any truncated expansion in $z$,
one is not able to fully reconstruct anyone of the $\sigma_k(\zeta)$'s.
Generally speaking, one expects the $1/n$-expansion to be more difficult
to implement than the $z$-expansion, but at the same time to contain a more rich dynamics.

For simplicity's sake, in order to illustrate the method avoiding asymptotic expansions of special functions,
let us analyze some one-dimensional barrier models.
We begin with the so-called Winter model, describing a particle on the half-line
$x\ge 0$, subjected to a potential spike at some point \cite{flugge,winter,primo,secondo,terzo}.
The Hamiltonian, after proper rescaling of space, time and energy, reads:
\beq
\label{Hwinter} 
\hat{H} \, = \, - \frac{\partial^2}{\partial x^2} \, - \, \frac{1}{\pi z} \, \delta(x-\pi) 
\eeq 
on the positive half line $x \ge 0$ and with vanishing boundary conditions at the origin: 
$\psi(x=0,t) \, = \, 0$.
We may also say that the model describes a particle contained inside a cavity --- the segment $(0,\pi)$
in our conventions --- with an impermeable wall at $x=0$ and a permeable one at $x=\pi$.
The model therefore implements the coupling of the box eigenfunctions to a set of continuum
states.
In the free limit $z\to 0$, the system decomposes into two non-interacting subsystems \cite{terzo}:
$(1)$ a particle in the box $(0,\pi)$, with a discrete spectrum only, composed of the
eigenfunctions
\beq
\sqrt{\frac{2}{\pi}} \sin(n x), \,\,\,\,\, n = 1,2,3,\cdots, \,\,\, 0 < x < \pi; 
\eeq
$(2)$ a particle on the half line $(\pi,\infty)$, having eigenfunctions 
in the continuum spectrum only, of the form:
\beq
\sqrt{\frac{2}{\pi}} \sin\big[k (x-\pi)\big], \,\,\,\,\,\, k > 0, \,\,\,\,\,\,\, \pi < x < \infty .
\eeq
As discussed in the Introduction, by switching on the coupling
from $z=0$ to $z\ne 0$, the discrete spectrum above turns into
an infinite tower of resonant states, having wavefunctions  are of the form:\footnote{
Due to the Gamow divergence \cite{gamow}, normalization is conventional 
(resonance wavefunctions are only locally integrable:
they are not even globally integrable in a weak sense).
}
\beq
\psi_n(x,t) \, = \,
\sqrt{ \frac{2}{\pi} } \, \exp\left[ - i \, E_n(z) \, t \right]
\left\{ 
\begin{array}{ccc} 
           \sin \left[ k_n(z) \, x \right]            & {\rm \,\, for} ~ 0\le x \le \pi ; 
\nonumber\\
a_n(z) \exp\left[ i \, k_n(z) \, x \right]  & {\rm \,\, for} ~ \pi < x <\infty . 
\end{array} 
\right.   
 \eeq
The outside amplitude has the explicit expression:
\beq
a_n(z) \, \equiv \, \frac{ \pi \, z \, k_n(z) }{ 2 \pi i \, z \,  k_n(z) \, + \, 1 } .
\eeq
The $n^\mathrm{th}$ resonance has the complex energy $E_n(z) \, = \, k_n(z)^2$.
The quantity
\beq
k \, = \, k_n(z)
\eeq
is the generalized, complex momentum of $\psi_n$ and it is the only non-trivial quantity
to compute.
In order to simplify the coefficients of the equation determining the allowed $k$'s,
it is convenient to set:
\beq 
w \, \equiv \, 2 \pi i \, k , 
\eeq
For a fixed coupling $z$ (fixed physics), the allowed $w$'s satisfy the implicit equation: 
\beq
\label{transeq}
e^w - z w - 1 \, = \, 0 .
\eeq
The transcendental character of the above equation in $w$ is related to the fact
that the model has an infinite tower of resonances, implying that it
must admit an infinite number of solutions: 
with an algebraic equation, one would obtain only finite-order multivaluedness.
The $1/n$ expansion is obtained by setting: 
\beq 
w \, = \, w_n(z) \, \equiv \, 2 \pi i \, n \, + \, \sigma(n,\zeta) 
\eeq 
where $n\ne 0$ is an integer and the effective coupling is defined as:\footnote{
The additional factor $2\pi i$ is inserted just for
practical convenience.} 
\beq 
\zeta \, \equiv \, 2 \pi i \, n \, z . 
\eeq  
To generate a recursion in $1/n$, eq.(\ref{transeq}) is conveniently rewritten as: 
\beq
\sigma \, = \, \ln\left[ 1 + \zeta + \frac{1}{2\pi i n} \, \zeta \sigma \right] .
\eeq
By formally expanding $\sigma(n,\zeta)$ in inverse powers of $2\pi i n$,
\beq 
\sigma(n,\zeta) \, = \, \sum_{k=0}^\infty \frac{ 1 }{ (2 \pi i n)^k } \, \sigma_k(\zeta) ,
\eeq
one obtains for the leading-order function: 
\beq
\label{sigma0}
\sigma_0(\zeta) \, = \, \ln_0\left( 1 + \zeta  \right) ,
\eeq
where by $\ln_0(z)$ we denote the principal branch of the complex logarithm of $z\ne 0$,  
i.e. the one with $- \pi < \arg_0(z) \le \pi$.
The leading-order $(lo)$ decay rate reads:
\beq 
\label{Gamma0}
\Gamma_n^{lo}(z) \, = \, \frac{n}{\pi} \, \ln\left[ 1 + \left(2 \pi z n\right)^2 \right],
\eeq
where we have assumed $z$ real (the physical case) and $\ln(x)$ is the real-analysis 
logarithm of $x$. 
By expanding eq.(\ref{Gamma0}) in powers of $z$, one obtains all the $1/n$-leading contributions
to $\Gamma_n(z)$, of the form $n^{2l+1} z^{2l}$, $l\ge 1$.
It is remarkable that the lowest-order perturbative result $\Gamma_n(z) \approx n^3 z^2$
is converted, via the resummation implied by the $1/n$-expansion, into the large-$n$ asymptotic behavior
$\Gamma_n(z) \approx n \ln\left(n |z|\right)$ \cite{terzo}.
Eq.(\ref{Gamma0}) actually represents an improvement of the leading result in eq.(244) of \cite{terzo}
($g\equiv -z$).
The first few corrections explicitly read:
\bea
\sigma_1(\zeta) &=& + \, \frac{\zeta}{ 1 + \zeta } \, \ln_0\left( 1 + \zeta  \right)  ;
\\
\sigma_2(\zeta) &=& - \, \frac{1}{2} \, \left(\frac{\zeta}{ 1 + \zeta }\right)^2 \, 
\ln_0\left( 1 + \zeta  \right) \Big[ \ln_0\left( 1 + \zeta \right) - 2 \Big] .
\eea
The evaluation of higher-order terms is straightforward.
By means of the simple third-order expansion above, one already obtains an excellent approximation
to $w_n(z)$, even for small $n$: for $z=-0.1$, for example, our expansion gives for $n=1$
a value of $w_n(z)$ with a relative difference with respect the exact value of about $6\times 10^{-5}$,
the error decreasing down to $2\times 10^{-6}$ for $n=5$.
The reason for the convergence even at $n=1$ may be that the expansion parameter
is actually $2\pi n$, rather than $n$.
As discussed in the Introduction, since $\sigma_0(\zeta)$ is a truly complex function, 
resonances already appear at lowest order in $1/n$, while in the standard perturbative treatment,
resonances appear only at second order in $z$ \cite{flugge,winter,refbase1,primo}.
By expanding in powers of $z$ the functions $\sigma_0(\zeta)$, $\sigma_1(\zeta)$ and $\sigma_2(\zeta)$
above, one explicitly obtains all the terms of the form $(n z)^l$, $z (n z)^l$, $z^2 (n z)^l$, $l\ge 1$ 
respectively.

As a more elaborate application of the $1/n$-expansion, let us investigate the
resonance structure of a system consisting of a particle on the real line subjected
to a double $\delta$-potential with general couplings $z_0$ and $z_+$.
By a proper shift and rescaling of the $x$-coordinate,
we can assume the $\delta$'s to be centered at $x=0$ and at $x=\pi$, 
so that the Hamiltonian reads: 
\beq
\label{doublewell} 
\hat{H} \, = \, - \frac{\partial^2}{\partial x^2}
\, - \, \frac{1}{\pi z_0} \delta(x) \, - \, \frac{1}{\pi z_+} \delta(x-\pi) . 
\eeq
Unlike previous case, the coordinate $x$ has range in the entire real line in this model.
We may also say that the Hamiltonian in eq.(\ref{doublewell}) describes a particle 
in a cavity --- the segment $(0,\pi)$ --- with permeable
walls on both sides, rather than just at one side as in the case of the
Winter model. The model therefore implements a coupling of the box
eigenfunctions to {\it two} sets of continuum states.
By introducing $w$ as in the previous case, the non-trivial part of the
evaluation of the $n^\mathrm{th}$ resonance wavefunction involves again the 
evaluation of this quantity, which satisfies in this case the equation:
\beq
e^w - \left( 1 + z_0 w \right) \left( 1 + z_+ w \right) \, = \, 0 ,
\eeq
The $1/n$ expansion is derived similarly to the case of the Winter model,
so we report the final results only. In lowest order we find:
\beq
\sigma_0\left(\zeta_0,\zeta_+\right) \, = \, \ln_0\big[ \left( 1 + \zeta_0 \right)\left( 1 + \zeta_+ \right) \big],
\eeq
and in next-to-leading order:
\beq
\sigma_1\left(\zeta_0,\zeta_+\right) \, = \, 
\left( \frac{\zeta_0}{1+\zeta_0} + \frac{\zeta_+}{1+\zeta_+} \right) 
\ln_0\left[ \left( 1 + \zeta_0 \right)\left( 1 + \zeta_+ \right) \right] ,
\eeq
where $\zeta_i \, \equiv \, 2 \pi i n z_i$ for  $i=0,+$.
Let us observe that the leading function is, for real $z_0$ and $z_+$, 
the sum of two Winter-model leading functions,
\beq
\sigma_0\left(\zeta_0,\zeta_+\right) \, = \, \sigma_0\left(\zeta_0\right) + \sigma_0\left(\zeta_+\right),
\eeq
implying that the leading-order decay rate is the sum of the corresponding Winter model
contributions (see eq.(\ref{Gamma0})):
\beq
\Gamma_n^{lo}\left(z_0,z_+\right) \, = \, \Gamma_n^{lo}\left(z_0\right) \, + \, \Gamma_n^{lo}\left(z_+\right) .  
\eeq
The physical interpretation is that the amplitude inside the cavity
flows through each one of the barriers, {\it as if} the other one was impermeable. 

\begin{figure}[ht]
\begin{center}
\includegraphics[width=0.5\textwidth]{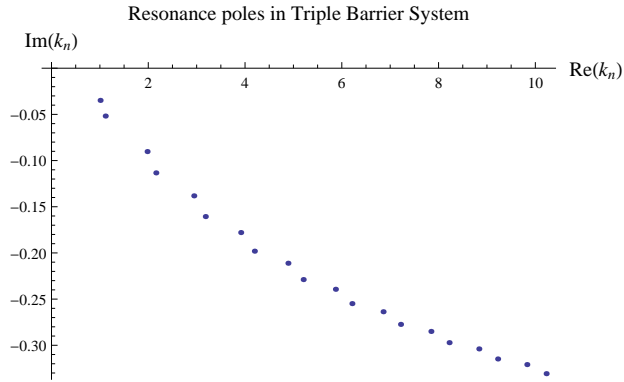}
\footnotesize
\caption{
\label{figura2}
\it Plot of the resonance poles $k_n = w_n/(2\pi i)$ in the fourth quadrant
of the complex $k$-plane for the triple-barrier model for $z_-=0.1$, $z_0=-0.05$, $z_+=0.15$.
The pairs of close poles correspond to the two determinations of the square root, i.e.
to $\sigma^+$ and $\sigma^-$ for the same $n$ (see text).
}
\end{center}
\end{figure}

As a final application of the $1/n$ expansion, let us consider the 3-$\delta$ model
with general couplings $z_-$, $z_0$ and $z_+$ and equal spacing between them, 
described by the Hamiltonian:
\beq
\label{triplewell} 
\hat{H} \, = \, - \frac{\partial^2}{\partial x^2} 
\, - \, \frac{1}{\pi z_-} \delta(x+\pi)
\, - \, \frac{1}{\pi z_0} \delta(x)
\, - \, \frac{1}{\pi z_+} \delta(x-\pi), 
\eeq
with the coordinate $x$ ranging in the real line. 
This model involves two cavities: the segments $(-\pi,0)$ and $(0,\pi)$, and two continuum 
sets of states, describing a particle lying in the half lines $(-\infty,-\pi)$ and $(\pi,\infty)$.
Each cavity is coupled to the other one for $z_0\ne 0$, as well as to the continuum states
on the same side of the real axis for $z_-, z_+ \ne 0$. 
Resonance wavevectors satisfy the transcendental equation:
\beq
\label{3delta}
a \, e^{2w} - b \, e^w + c \, = \, 0 ,
\eeq
where $a$, $b$ and $c$ are the following functions of $z_-, z_0, z_+$ and $w$:
\bea
a &\equiv& 1 - z_0 w ; \,\,\,\,\,\,\,\,\,\,\,\,\,\, b \, \equiv \,  2 + \left( z_- + z_+ \right) w  ; 
\nonumber\\
c &\equiv & 1 + \left( z_- + z_0 + z_+ \right) w
+ \left( z_- z_0 + z_- z_+ + z_0 z_+ \right) w^2 + z_- z_0 z_+ w^3  .
\eea
In eq.(\ref{3delta}), powers of $w$ up to the third one included are involved,
each multiplied by a symmetric polynomial in $z_-$, $z_0$ and $z_+$ of the same degree.
Unlike previous cases, where only the exponential term $e^w$ was involved, two exponential terms, $e^w$ and $e^{2w}$, 
are present in this case; we will see later that this fact is the "analytic source" of the degeneracy in the
resonance spectrum for $z_-=z_+$ in the limit $z_0\to 0$.
In order to generate the $1/n$-expansion in this case, the idea is to treat $e^w$ and $e^{2w}$ as unknown quantities, 
while treating the terms containing powers of $w$ as known quantities. By setting
$\xi \, = \, e^w$, with $w \, = \, 2 \pi i n + \sigma^\pm$,
we obtain the implicit equation on $\sigma^\pm$:
\beq
\sigma^\pm \, = \, \log_0
\left\{ 
\frac{ 2 + \left( 2 \pi i n + \sigma^\pm \right) 
\left[
z_+ + z_- \pm H\left(z_-,z_0,z_+,\sigma^\pm\right)
 \right] }
{2\left[ 1 - ( 2 \pi i n + \sigma^\pm) z_0 \right]} 
\right\} ,
\eeq
where we have defined:
\beq
H\left(z_-,z_0,z_+,\sigma^\pm\right) \, \equiv \, \Big\{ \left(z_+ - z_-\right)^2 + 4 z_0^2 \, \Big[1 + \left( 2 \pi i n + \sigma^\pm\right)z_+\Big]
\Big[ 1 + \left( 2 \pi i n + \sigma^\pm\right)z_- \Big] \Big\}^{1/2} .
\eeq
For the determination of $z^{1/2}$, let us take the principal branch, i.e. the one with $-\pi<\arg(z) \le \pi$.
The equation above is solved recursively by setting as usual:
\beq
\sigma^\pm\left(n; \zeta_-, \zeta_0, \zeta_+ \right) \, = \, 
\sum_{k=0}^\infty  \frac{1}{(2 \pi i n)^k} \, \sigma_k^\pm\left(\zeta_-, \zeta_0, \zeta_+ \right) ,
\eeq
where $\zeta_i \, \equiv \, 2 \pi i n z_i$ for $i = -, 0, + $.
At the lowest order in $1/n$, we obtain the rather compact formula:
\beq
\label{triplelo}
\sigma_0^\pm\left( \zeta_-, \zeta_0, \zeta_+  \right) \, = \, \log_0 
\left[
\frac{ 2 + \zeta_+ + \zeta_- \pm \Delta\left(\zeta_-,\zeta_0,\zeta_+\right)   }{2(1-\zeta_0)}
\right] ,
\eeq
where:
\beq
\Delta\left(\zeta_-,\zeta_0,\zeta_+\right) \, \equiv \, 
\big[ \left(\zeta_+ - \zeta_-\right)^2 + 4 \, {\zeta_0}^2 \, (1+\zeta_+)(1+\zeta_-) \big]^{1/2} .
\eeq
By means of the Mathematica system \cite{mathematica}, the next-to-leading-order 
correction is obtained:   
\beq
\label{triplenlo}
\sigma_1^\pm\left( \zeta_-, \zeta_0, \zeta_+  \right) \, = \,  
\frac{N^\pm\left( \zeta_-, \zeta_0, \zeta_+  \right)}{D^\pm\left( \zeta_-, \zeta_0, \zeta_+  \right)} \, \log_0 
\left[
\frac{ 2 + \zeta_+ + \zeta_- \pm \Delta\left(\zeta_-,\zeta_0,\zeta_+\right)   }{2(1-\zeta_0)}
\right] \,  ,
\eeq
where $N^\pm$, $D^\pm$ are the following algebraic functions:
\bea
N^\pm\left( \zeta_-, \zeta_0, \zeta_+  \right) &\equiv& - 2 \zeta_0^3 \left( \zeta_- + \zeta_+ + 2 \zeta_-\zeta_+ \right) 
+ \zeta_0^2 \left( 4 + 6 \zeta_- + 6 \zeta_+ + 8 \zeta_-\zeta_+ \right) 
+ \left(\zeta_- - \zeta_+\right)^2 +
\nonumber\\
&&\,\,\,\,\,\,\,\,\,\,\,\,\,\,\,\,\,\,\,\,\,\,\,\,\,\,\,\,\,\,\,\,\,\,\,\,\,\,\,\,\,\,
\,\,\,\,\,\,\,\,\,\,\,\,\,\, \,\,\,\,\,\,\,\,\,\,\,\,\,\,\,\,\,\,\,\,\, 
\pm \left(2\zeta_0+\zeta_-+\zeta_+\right)\Delta\left(\zeta_-,\zeta_0,\zeta_+\right) ,
\eea
and
\beq
D^\pm\left( \zeta_-, \zeta_0, \zeta_+  \right) \equiv \left(1-\zeta_0\right) \Big[ (\zeta_- - \zeta_+)^2  + 4 {\zeta_0}^2 (1+\zeta_+)(1+\zeta_-)
\pm (2 + \zeta_- + \zeta_+) \Delta\left( \zeta_-, \zeta_0, \zeta_+  \right) \Big] .
\eeq
As expected on the basis of physical intuition, eqs.(\ref{triplelo}) and (\ref{triplenlo}) exhibit resonance degeneracy 
for $\zeta_0\to 0$ and $\zeta_-=\zeta_+$. 
The expressions above largely simplify in particular cases, such as for example equal couplings of the
side barriers, $\zeta_-=\zeta_+$, or small coupling of the intermediate barrier, $|\zeta_0|\ll |\zeta_-|,|\zeta_+|$.
A plot of the resonance poles, in the next-to-leading approximation specified by the $\sigma_0^\pm$ and $\sigma_1^\pm$
above, is given in fig.\ref{figura2}.

Let us conclude by saying that the calculation of resonant state parameters, such as energy distributions, 
decay widths, branching ratios, etc., is of primary importance in applications of non-relativistic quantum mechanics
\cite{refbase1,refbase2}. 
This problem is generally treated, apart from the simplest cases, with specific numerical methods \cite{hatano}. 
We have presented in this letter a general analytic expansion in $1/n$ for resonance states,
where $n=1,2,3,\cdots$ is the excitation number, which substantially simplifies the relevant equations.
In order to avoid asymptotic expansions of special functions and obtain compact analytic formulas,
we applied the $1/n$-expansion to some simple one-dimensional barrier systems, namely particles subjected 
to $\delta$-like potentials. 
Let us remark however that the expansion presented is quite general
and can be applied to any systems containing an infinite tower of resonances.
On the mathematical side, the $1/n$ expansion can also be applied to solve transcendental equations involving 
exponential and power functions of the unknown variable.

Expansions of similar form as the one presented in this paper are well known in statistical physics and 
quantum field theory \cite{coleman},
the main difference being that in the latter cases the parameter $n\gg 1$ denotes the number
of components of a field or the order of some symmetry group, i.e. a state-independent quantity
entering the Hamiltonian of the system.
The idea of an approximate, all-order, resummation of the perturbative series --- as opposed
to an exact, fixed-order, perturbative calculation --- is basic in the studies of the current 
theory of strong interactions, the so-called Quantum-Chromodynamics (QCD). 
In the latter case, the role of our small quantity $z$ is played by the coupling constant
at a large momentum transfer $Q$, $\alpha_S(Q) \ll 1$, while the role of the large parameter $n$
is played by the square of a large logarithm $L^2$ of infrared (soft and collinear) origin,
or by a large logarithm $L$ of infrared or ultraviolet origin.

\vskip 0.4 truecm

\centerline{\bf Acknowledgments}

\vskip 0.4 truecm

\noindent
I would like to thank M. Bochicchio for discussions.

\end{document}